# Indirect Methods in Nuclear Astrophysics

Livius Trache, Florin Carstoiu

*"Horia Hulubei" – National Institute for Physics and Nuclear Engineering*
*Bucharest-Magurele, str. Reactorului 30, RO-077125, Romania*

Abstract

This paper follows the inaugural talk one of the authors (LT) gave at the opening of the ECT* workshop with the same title, which he co-organized in Trento, Italy, November 5-9, 2018. As such it follows the ideas expressed there, which were to outline the discussions that the organizers intended for that meeting. Therefore, the paper will review the indirect methods in nuclear astrophysics, their use and their specific problems, old and new, the need to further developments rather than giving complete treatments of each method or reviewing exhaustively the existing literature. The workshop was from its inception aiming also at reviewing the status of the field of nuclear astrophysics and its connections with adjacent branches of physics. Some lines on these are included here.

**Keywords:** Nuclear astrophysics, nuclear physics for astrophysics, indirect methods, Coulomb dissociation, one-nucleon transfer, nuclear breakup, resonances, Trojan Horse method, decay spectroscopy

**PACS numbers:** 26.50.+x; 25.60.-t; 25.60.Je; 25.60.Gc

## 1 Introduction

As announced in the abstract, this paper follows the introductory remarks given at the ECT* workshop "Indirect Methods in Nuclear Astrophysics" (IMNA), Trento, Italy, November 5-9, 2018 [1]. The remarks announced the main topics, how the organizers setup the invited speakers list and the way they conceived the progress of the lectures and of the discussions.

We start from the premise that nuclear astrophysics (NA) is in the last few decades an important part of the science programs of all nuclear physics laboratories. Moreover, especially in a time when the concept of multi-messenger observations becomes not only used and validated by the scientific community, but widely known to the larger public, nuclear astrophysics must be redefined to include (or being close to):

- Nuclear physics for astrophysics (NPA)

- Stellar dynamics
- Nucleosynthesis modelling
- (specific) astrophysics observations: X-ray and Gamma-ray space telescopes, cosmochemistry.

Even Cosmology – a very large field in itself - becomes closer to NA and there are mutual benefits.

These said, there is clearly a need for closer interaction among the specialists in these fields. It is obvious from the title of the workshop that the focus was intended to be nuclear physics for astrophysics, but we appealed to specialists in the adjacent sub-fields listed above to come and talk about the progress on specific topics of interest and in particular about their needs for new or more precise nuclear data. Similarly, we wanted them to listen to talks about the current possibilities, limitations and problems of the nuclear physicists working with indirect methods for nuclear astrophysics.

As such the main topics in Nuclear Physics for Astrophysics, related to indirect methods to be discussed were:
- Nuclear astrophysics for practitioners, basics. Nuclear data needs;
- Stellar dynamics, nucleosynthesis modeling, observations;
- Review of existing indirect methods in nuclear astrophysics:
  - "the list";
  - Specifics. Assessment of problems with the accuracy of each indirect methods, experimental and theoretical, the importance of calculated absolute values;
  - The need for modern theories and codes; parameters to use in calculations;
- Review of experimental methods, equipment and specifics;
- New facilities, including RIB facilities, and their nuclear astrophysics programs;
- Related topics – new directions.

While as announced, the paper is based on the inaugural ECT* talk, in a few cases we will add figures to illustrate better the indirect methods described. In most of the cases we will use illustrations from own work, for reasons easy to understand, some already published or shown at past conferences. We felt this need for illustrations because presumably the readership for this paper is wider and less knowledgeable in this topic than the workshop's attendees. We will insist more on giving general descriptions of the basic ideas of the methods, their potential and of the needs for improvements, rather than exhaustive discussions and examples. Will send the interested reader to suitable literature.

The paper is structured as follows: after this Introduction, in Sect. 2 a few general considerations on NPA methods are given to set up the framework and substantiate the later discussions. In Sect. 3 after a short discussion on the paradigm used in IMNA, a list of indirect methods is given, while each method is briefly discussed in the subsections of Sect. 4. Section 5 closes with discussions on a few adjacent topics of nuclear astrophysics and some conclusions.

## 2. Nuclear physics for astrophysics

We know for about a century that nuclear reactions are the fuel of the stars (Edington, 1920) and the origin of chemical elements in the Universe. This latter through phenomena called globally nucleosyn-thesis that took place both in the Big Bang [2] and later in stars (see Burbidge, Burbidge, Hoyle & Fowler [3] and Cameron [4], both 1957). We also know that it continues today (see, e.g. [5] or think about the Sun [6]). There is no doubt about these in the scientific community, which means that there are proofs or many arguments for these statements! All these proofs are based on astrophysics observations and on quantitative modelling of nucleosynthesis, using nuclear data. However, we are far from understanding fully nucleosynthesis, to know the places where various processes have happened, or from having a good quantitative description for them, etc. These are subjects that nuclear astrophysics deals with.

We will not treat here the basics of nuclear astrophysics, will not introduce concepts like the astrophysical S-factors, reaction rates, Gamow window and will rather refer the reader to many introductory texts, like [7], e.g., if necessary.

Nuclear Physics for Astrophysics in particular, aims at providing data and models for the understanding of the origin of chemical elements in the Universe. We do not have a complete quantitative explanation of the creation of all elements, despite many successes in the last decades.

There are two types of experiments in nuclear physics for astrophysics:
- *Direct measurements*, that is to reproduce and measure in the nuclear physics laboratory the reactions that happened or happen in stars, at exactly those relevant energies (in the Gamow window). The latter is a big problem, as the "stars are cold" on nuclear physics' scale of energies, and for reactions between charged particles the Coulomb barrier leads to very small cross sections and these experiments are difficult. At low energies the signal-to-background ratio becomes very small and special measures must be taken to improve it. In most cases the data had to be extrapolated down to energies in the Gamow window. Progress was made and will continue using underground laboratories, existing (LUNA at Grand Sasso National Laboratory) or planned (in USA, China, etc.). It is not that far back in time that the first measurements in the Gamow window were made [8], avoiding uncertain extrapolations.
- *Indirect measurements:* experiments are done using beams at nuclear laboratory energies of 1s, 10s, 100s MeV/nucleon to extract data to be used for the evaluation of cross sections at energies of 1s, 10s, 100s of keV/nucleon, relevant in stars. There are two main reasons we must resort to indirect methods in NA:
  - The very low cross sections mentioned above when we attempt reactions at energies relevant in stars (1s-100s keV).
  - Many, in fact most, of the reactions occurring in different NS processes involve unstable nuclei. Therefore, we need to use radioactive species for the experimental determination of needed nuclear data. Only a few experiments could be done with radioactive targets for situations where the nuclides involved have a reasonably long lifetime and can be produced ($^7$Be [9], $^{22}$Na [10], to mention only the pioneering ones), but mostly we use radioactive ion beams (RIB). Moreover, as the direct measurements are very difficult even with stable nuclei, as said before, due to the very low cross sections, measurements at low energies with unstable species are out of experimentalists' reach for now (there are pioneering attempts with decelerated beams at GSI [11], though, and soon at other places). Therefore, most of the reactions involving unstable nuclei are being studied using *indirect methods*. We shall review these methods here.

There are tens of thousands of nuclear reactions and nuclear processes that occur in stars. Some are very important, some are less important, and some may be irrelevant in one type of process, while becoming important in another, depending on the conditions of the particular process and environment: composition, densities and temperatures involved. There are also many nucleosynthesis processes, and our knowledge about them differs. To have an evaluation of which data are of importance and in which circumstances, to what precision they are necessary for good, reliable, nucleosynthesis modelling, is very important for those of us working in obtaining data for nuclear astrophysics. It is a crucial point and on its importance we insisted at the workshop but will not discuss here.

## 3. Indirect Methods in Nuclear Astrophysics. The list.

Before going to make the list of the indirect methods for nuclear astrophysics, we have to say that it has to be, by necessity, a personal view of the current list. The methods can be organized and certainly ordered differently than below. We would start by saying that the first indirect nuclear data that were used for astrophysics were the mass measurements of the early XX-th century. Those measurements and the $E=mc^2$ of A. Einstein lead Sir A. Edington to suppose that solar energy arises from nuclear reactions. Further, the beta-decay studies allowed Critchfield and Bethe to propose and evaluate the pp-chain of reactions [12]. Then the lack of knowledge on the mass gaps at A=5 and A=8 lead to the wrong, but historically important, model of nucleosynthesis by Alpher, Bethe and Gamow published on April 1, 1948 [13]. And one could go on and on!

However, explicit proposals for indirect methods in nuclear astrophysics started gaining momentum in the mid-eighties of last century.

The list of IMNA, as presented and discussed at the workshop is:
  A. Coulomb Dissociation
  B. Single-nucleon transfer reactions – the ANC method
  C. Nuclear breakup reactions
  D. The Trojan Horse Method
  E. Spectroscopy of resonances, a wide category of reactions, types of experiments and theories.

While it is clear that the indirect methods in the list above may differ from one another by laboratory energies at which they are applied and by the techniques used, both experimental and theoretical, there is a common path from their results to the evaluation of the cross sections or reaction rates at energies or temperatures relevant in stellar processes:
  1. Experiments are made at energies usual for the nuclear physics laboratories
  2. Theoretical (reaction) calculations are made
  3. The experimental results of step (1) are compared with calculations of step (2) to extract nuclear information (typically nuclear structure parameters)
  4. The extracted information is used to evaluate nuclear astrophysics data: cross sections, astrophysical S-factors or reaction rates.

These steps are also sketched in Figure 1. There (B on the right-hand side) a point is also made to show that in both steps 2 and 4, additional knowledge is very important. The theories and the parame-ters used in both situations, at large and at low energies, need to be well established and vetted throughout in order to give confidence on the end results. We should stress that most of the time it is important to have good, reliable calculations of the absolute values at point (2), a feature not exactly common to nuclear reaction theories. Another point that is not figured there and not specified above is the importance of the choice of the data/information that we extract at point (3), their relevance for the precision of the evaluation at step (4), and in particular the need that they are model independent, as much as possible. This will be exemplified when the ANC method Bill be discussed.

Another important step is to compare the results of the indirect methods (step 4) with results of direct measurements, if they exist (A, on the left side of the figure).

We shall start discussing them briefly in list's order.

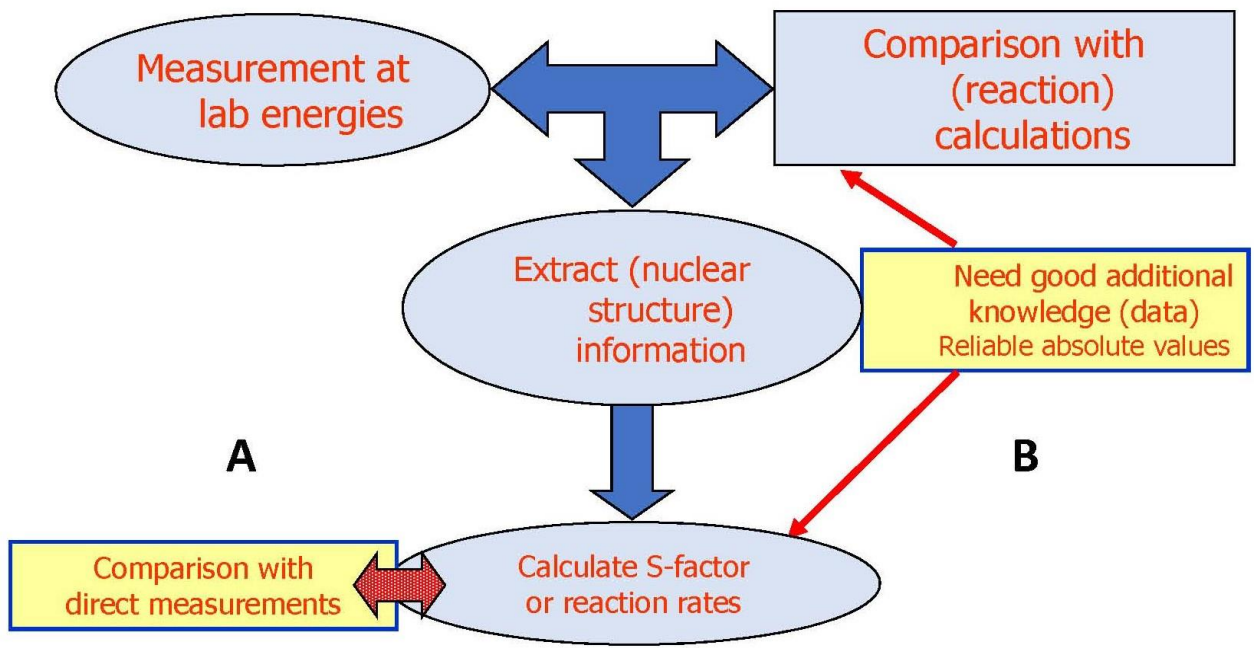

**Figure 1. The paradigm used in Indirect Methods in Nuclear Astrophysics.**

## 4. Indirect Methods in Nuclear Astrophysics

### A. The Coulomb dissociation

The Coulomb dissociation is a method specifically introduced for nuclear astrophysics over thirty years ago. Schematically, it works as follows.
- Instead of studying the **radiative proton capture reaction** X(p,γ)Y at a definite center-of-mass energy $E_p$, a process in which a gamma-ray of energy $E_\gamma = E_p + S_p$ is emitted ($S_p$=binding energy of the proton in nucleus Y) after the capture of a proton, we could measure the inverse process: **photodissociation**. A photon of energy $E_\gamma$ produces the dissociation Y + γ → X + p, in which a proton-core system of relative energy $E_p = E_\gamma - S_p$ results. Then the Fermi golden rule of detailed balance can be used to relate the cross section of the two processes. Obviously, the energy of the photon involved must be larger than the binding energy $S_p$.
- Baur, Bertulani and Rebel [14] proposed **to replace the real photons** needed in **photodissociation with virtual photons**. A fast-moving projectile Y in the Coulomb field of a target senses a field of virtual photons that induces the dissociation of the projectile Y → X + p. The resulting cross section for Coulomb dissociation is a product between the photodissociation cross section and the number of virtual photons of each multipolarity and energy needed:

$$\frac{d^2\sigma}{dE_\gamma d\Omega}(E_\gamma,\theta) = \frac{1}{E_\gamma}\left[\frac{dN(E1,E_\gamma)}{d\Omega}\sigma_{E1}^{photo}(E_\gamma) + \frac{dN(E2,E_\gamma)}{d\Omega}\sigma_{E2}^{photo}(E_\gamma) + ...\right]$$

To increase the effect, a strong Coulomb field of a high Z target is needed (Pb for example). Only the photons with energies higher than $S_p$ contribute in the dissociation and a continuum spectrum of relative energies $E_p$ is obtained. The photodissociation cross section is then directly related to the radiative capture cross section sought in nuclear astrophysics.

Problems arise from:
- the need of relatively large projectile incident energies to produce enough virtual protons of the large energy $E_\gamma > S_p$ necessary to produce photodissociation. This condition is easily satisfied by the new RIB facilities.
- the fact that different multipoles do contribute in different proportions in Coulomb dissociation and in radiative capture (see eq. above). Therefore, the disentangling of different multipole contributions from angular distribution measurements in Coulomb dissociation is needed before transforming the results into astrophysical S-factors for radiative capture. That is experimentally very demanding at the large projectile energies necessary to satisfy the first condition. Mostly one relies on calculations so far, but setups are conceived currently to resolve this experimentally [15].
- the difficulty to separate the contribution of the nuclear and Coulomb fields in dissociation at large energies. This is done selecting dissociations that happen at large impact parameters, which translates into measurements very close to zero degrees, experimentally a very difficult task. The problem is further complicated by the (usually) poor definition of the currently available radioactive beams.

However, a large number of very good Coulomb experiments have been done so far to obtain astrophysical data, and the method is considered rather well established [16]. One important conceptual advantage of the method is that from Coulomb dissociation the energy dependence of the astrophysical S-factor $S(E_p)$ can be experimentally extracted ($E_p$ = p-core relative energy). While experimental difficulties restrict measurements very close to the threshold, that is at the equivalent of capture energies in the Gamow window, and therefore one needs again extrapolations, the measurement of the excitation function $S(E)$ may also give information about the location and widths of low energy resonances of potential importance in nuclear astrophysics.

Currently the method is considered appropriate for use with proton rich radioactive beams at intermediate energies obtained through projectile fragmentation. However, there are no principle restrictions to use it for radiative alpha capture reactions (attempts were made and further ones are planned to study the Coulomb dissociation of $^{16}O \rightarrow \alpha + ^{12}C$, the inverse of the $^{12}C(\alpha,\gamma)^{16}O$ reaction).

The method needs further improvements in experiments, in particular in the multipole decomposition, while in theory further attention must be given to the interference with the nuclear component. It is clear that Coulomb dissociation will remain an important tool for nuclear astrophysics in the era of radioactive ion beams.

## B. Single-nucleon transfer reactions – the ANC method

A direct reaction is characterized by the involvement of a limited number of degrees of freedom, or the rearrangement of one or of a few nucleons during a fast process. From the early days of nuclear physics, nucleon transfer reactions were the way to study the single-particle degrees of freedom of nuclei. Typically, spectra of final states and angular distributions are measured. Due to the direct character of the interaction, the tool of choice for the description of transfer reactions is the Born Approximation, either in the Plane Wave (PWBA), or the Distorted Wave (DWBA) form:

- by comparing the shape of the measured angular distributions with DWBA calculations, the quantum numbers *nlj* of the single-particle orbitals involved could be determined (not always uniquely), and
- by comparing the absolute values of experimental cross sections with those calculated, the spectroscopic factors $S_{nlj}$ can be determined for the states populated.

The spectroscopic factor is proportional to the "probability" that a many-body system (the nucleus) is found in a certain configuration. In the case we are talking about, transfer of one nucleon to/from a single particle orbital with quantum numbers *nlj*, the classical definition (from Macfarlane and French, 1960 to Bohr and Mottelson, 1969) relates the spectroscopic factors *S(nlj)* to the occupation number for the *nlj* orbital in question. One nuclear state may present several spectroscopic factors due to configuration mixing: e.g. the ground state (g.s.) of $^8$B has $S(p_{3/2})$ and $S(p_{1/2})$, related to the probability that the last proton is bound around the g.s. of the $^7$Be core in a $1p_{3/2}$ and a $1p_{1/2}$ orbital. The determination of spectroscopic factors from one-nucleon transfer reactions was and is crucial in building our current understanding of the fermionic degrees of freedom in nuclei and their coupling to other types of excitations. However, in determining the absolute values of the spectroscopic factors as the ratio between the experimental cross section and the DWBA calculated cross section one makes (1) a strong assumption that the single-particle configuration assumed is dominant in the wave function (actually in the contribution to the cross section measured) of the state under consideration and (2) that the parameters used in the DWBA calculations are appropriate.

A connection between transfer reactions and nuclear astrophysics was made in the 1970s by Claus Rolfs [17] but in the opposite direction (NA gives info about nuclear spectroscopy). The Asymptotic Normalization Coefficient (ANC) method is an indirect NA method introduced systematically by the Texas A&M group and successfully and extensively used to determine astrophysical S-factors for the non-resonant component of radiative proton capture at low energies (zero to tens or hundreds of keV) from one-proton transfer reactions involving complex nuclei at laboratory energies (about 10 MeV/u) [18-20]. The method was explained in detail in many previous publications, we summarize the main ideas below and in Figure 2, taken from Ref. 21. Essentially it works around the problem of the considerable dependence of the absolute values of the extracted spectroscopic factors on the parameters used in the DWBA calculations. It works for cases where the transfer reactions are peripheral, a condition that may be fulfilled by choosing the target-projectile combinations and the bombarding energies. We shall go briefly through the basics of the method.

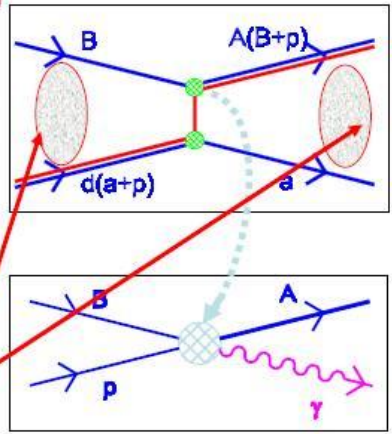

**Figure 2. The Asymptotic Normalization Coefficient method compactly explained (see text).**

We use peripheral proton transfer reactions to extract the ANCs, which can be used to evaluate (p,γ) cross sections important in different types of H-burning processes. The idea behind it is that in peripheral processes it is sufficient to know the radial wave functions at large distances, and this asymptotic radial behavior is given by a known Whittaker function times a normalization coefficient $C_{nlj}$ (this is the asymptotic normalization constant, or ANC, as in the equation on the lower right corner of Fig. 2). That allows the evaluation of the overlap integrals $I$ which enter in the DWBA calculations (first equation in Figure 2) and from there $C_{nlj}$ can be determined by comparison with the experiment. In the transfer reaction B(d,a)A one has to know one of the two vertices (the spectroscopic factor $S_i$ or the ANC for one vertex, the lower in the top diagram on right) to determine the spectroscopic factor $S_f$ or the ANC for the other one. And from there one can calculate the radiative capture cross section for the B(p,γ)A process (lower diagram on right) as it is only sensitive to the peripheral behavior of the overlap integral. The quantities $b_{nlj}$ are the single-particle ANC, that is, the asymptotic coefficients for the radial functions normalized to unity. These are those used in the DWBA calculations. It has been shown that extracting the ANC is less parameter dependent than extracting spectroscopic factors (see Fig. 11 in [18], e.g.). The parameters varied here are those defining the geometry of the Woods-Saxon well that binds the proton around the core: the reduced radius $r_0$ and the diffuseness $a$. Figure 2 also stresses the importance of having good and reliable optical model potentials (OMP) to make the DWBA calculations, a problem we will discuss later here. Note: the independence of the ANC extracted from the parameters of the Woods-Saxon potentials used to calculate the radial wave functions above should not be confused with an independence on the parameters of the typically Woods-Saxon shaped optical model potentials used to calculate the distorted wave functions of the scattering! Good care should be taken to extract or evaluate good OMP in both the entrance and exit channels of the reaction. The absolute values depend very much on these OMP parameters, in most cases more than on the $(r_0,a)$ parameters of the proton-binding potential well.

The ANC method was used in several experiments of this type. We will show a typical one of the studies, on the $^{12}$N(p,γ)$^{13}$O proton capture reaction at stellar energies. It uses the proton transfer reaction $^{14}$N($^{12}$N,$^{13}$O)$^{13}$C with a $^{12}$N beam at 12 MeV/u [22]. Figure 3 below, also the image of a slide shown during a lecture on the subject, summarizes the whole process, from extracting the data from experiment to nuclear astrophysics conclusions. Going from bottom left, clockwise:
- we have measured the elastic scattering and the one-proton transfer using a $^{12}$N beam produced and separated with the MARS spectrometer [23] at Texas A&M University. The elastic scattering data (lower left corner) were used to determine the OMP needed in the DWBA calculations for transfer.
- The ANC for the system $^{13}$O→$^{12}$N+p was extracted from the transfer data (top left) after which
- the ANC was used to evaluate the non-resonant component of the astrophysical S-factor for the radiative proton capture $^{12}$N(p,γ)$^{13}$O and the corresponding reaction rate as a function of stellar temperature (top right).
- Finally, the astrophysical consequences are shown in a plot (bottom right) which shows the region of density-temperature where the capture process competes with its competitor (β-decay), in first stars (above the full line 1). For comparison, the curves from literature before our data were measured are shown. There is a big change from the original estimates (dashed curves) based on theoretical estimates only, showing the importance of experimental measurements.

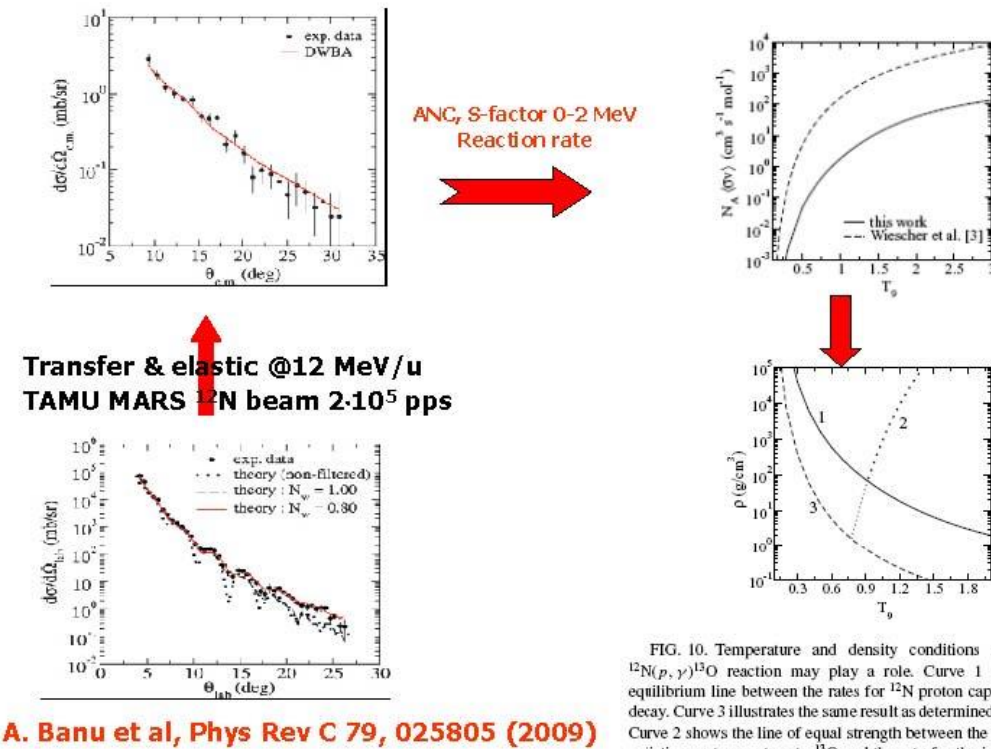

**Figure 3.** Summary of how elastic and one-proton transfer data measured with secondary RIB (clockwise from lower left side along the arrows) are transformed in nuclear astrophysics information (bottom right side) (from Ref. 21).

A variation of the ANC method uses one-neutron transfer reactions to obtain information about the mirror nuclei, for example studying the $^{13}C(^{7}Li,^{8}Li)^{12}C$ reaction to determine the ANC for $^{8}Li$ which one then translates into the corresponding structure information (the proton ANC) for its mirror $^{8}B$ and from there $S_{17}(0)$ for the reaction important in the neutrino production in Sun $^{7}Be(p,\gamma)^{8}B$ [6]. We did this using the mirror symmetry of these nuclei: the similarity of their wave functions, expressed best by the identity of the neutron and proton spectroscopic factors for the same *nlj* orbital in the two nuclei $S_p(nlj)=S_n(nlj)$ (of course, the radial wave functions are not identical!). The experiment using these concepts and the results were published in Ref. 24.

We mentioned before that in order to extract data, either the spectroscopic factors, or the ANCs, the experiments must be compared with calculations, and in the above conditions, the knowledge of the optical potentials is crucial. This is an important problem, with no clear solution so far and on which we need better data and better theories and codes. There is not only the usual problem that we know from elastic scattering data that the OMP extracted for nucleus-nucleus interactions are not unique, but the current quality of elastic scattering data with radioactive beams is not sufficient to extract good OMP. In order to avoid the ambiguities usually related to the fits of the elastic scattering data, better data are needed, especially data extended to larger angular ranges, including data at backward angles, where the cross sections become very small. There are several attempts to establish procedures leading to reliable predictions for optical potentials, none globally accepted. Certainly, more work is needed in this direction: experiments, systematics and theoretical analyses. We want to draw the attention here that we do not only need 'new data' for this purpose, but 'better data' as well, in order to advance on this topic, of crucial importance in many types of experiments involving radioactive beams. The matter was discussed at the workshop, but the only conclusion was "more work is needed, better data are needed". This may imply better and more precise data possible only with stable beams. A proposed line of work is that of our group that has established a procedure based on double folding, starting from an effective nucleon-nucleon interaction called JLM and many successes were obtained with it. We will not insist on this here, but we send you to literature [25].

We conclude that while the experimental conditions when using transfer reactions with RIBs need improvement, resolutions in particular, angular and energy resolutions, further work is due also in theory. Not only the improvement on OMP, but on reaction theories, codes and parameters. Careful evaluation of the improvements brought in by the increased use of extended calculations, like the use of coupled channels discrete calculations (CCDC) must be discussed and assessed.

## C. Nuclear breakup reactions

After the discovery of the first halo nucleus $^{11}$Li [26], much work was done for the study of radioactive beams, and in particular of loosely bound nuclei. Several laboratories have demonstrated that one-nucleon removal reactions (or breakup reactions) can be a good and reliable spectroscopic tool for such nuclei. In a typical experiment a loosely bound projectile at energies above the Fermi energy impinges on a target and loses one nucleon. The momentum distributions (parallel and/or transversal) of the remaining core measured after reaction were relatively easy to measure and they gave information about the momentum distribution of the removed nucleon in the wave function of the ground state of the projectile. The shape of the distributions was shown to be sensitive to the quantum numbers *nlj* of the single particle wave function (determining unambiguously only the orbital angular momentum *l*; shell model systematics are needed for the others) and in some cases even to assess the mixing of different configurations in the ground state wave function of the projectile (see Sauvan [27] for example). At later stages coincidences between the cores and gamma-rays allowed even for the determination of complex configuration mixings. Most of the cases studied involved neutron removal reactions on light targets like Be or C, where the nuclear breakup dominates. The method is also valuable because it can be applied using low quality radioactive ion beams available so far: low intensities, down to a few pps, and poor definition (energy and direction resolutions). Typically, these beams were/are from fragmentation reactions and the energies for which the technique is applicable must be above the Fermi energy in nuclei, intermediate energies E>25-50 AMeV (which is always the case for fragmentation).

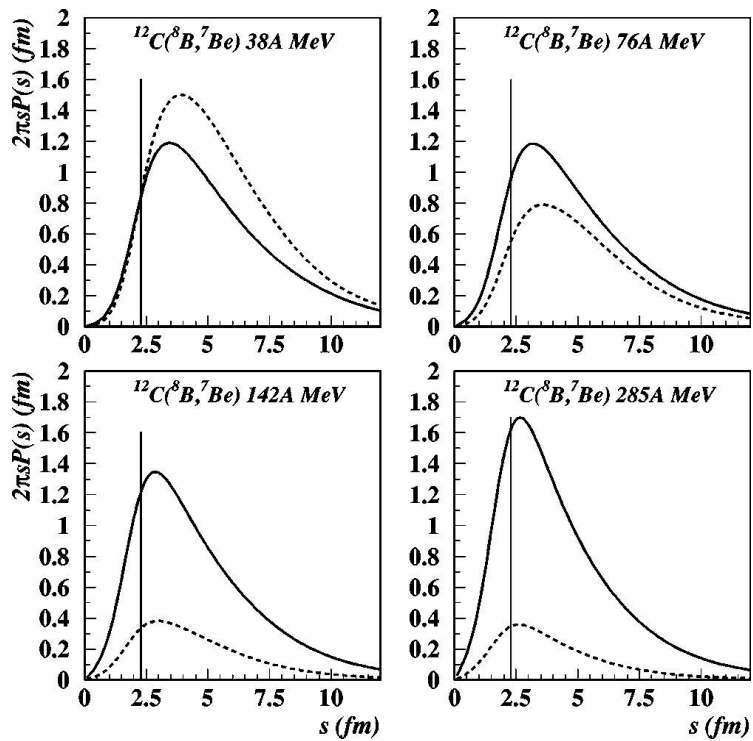

Figure 4. The breakup probability profile as a function of the impact parameter for the case of $^8$B→$^7$Be+p on a light target at various energies. The vertical line shows the position of the $^7$Be core rms radius. The stripping (full lines) and diffraction dissociation (dashed lines) components are shown. From Ref. 29.

Later it was shown in Ref. 28 that on a large range of projectile energies breakup reactions are peripheral (Figure 3) and, therefore, the breakup cross sections can be used to extract asymptotic normalization coefficients. In cases where one proton removal Y→X+ p is studied, the ANC found can be used to evaluate the corresponding radiative proton capture cross sections X(p,γ)Y at very low energies, useful in NA. For these to be correct, on one hand one must measure the absolute values of the breakup cross sections and to have reliable reaction model calculations and tested codes and parameters, on the other hand. The calculations must reproduce the available data from measurements in order to be tested. This is a very important point, which we stressed in the workshop. The method to use breakup reaction for nuclear astrophysics was first applied in [28, 29] to the breakup of $^8$B to determine $S_{17}(0)$. It was shown that all available breakup data, on targets from C to Pb and at energies from 27 MeV/u to 1400 MeV/u lead to a consistent value for the ANC for $^8$B→$^7$Be+p. Different reaction models and different nucleon-nucleon effective interactions were used. The overall uncertainty estimated at about 10%, which is a very good agreement, a fact that validated both the $S_{17}(0)$ adopted in the neutrino production calculations pertinent to what was called the "solar neutrino puzzle" before the neutrino oscillations were demonstrated, and the validity of this indirect method in NA.

One other example useful to mention here is the breakup of $^{23}$Al at intermediate energies. It is a more complex experimental situation where several configurations contribute to make the ground state of the $^{23}$Al projectile. The ground state of this projectile has configuration mixing (it turned out to be 4 different configurations). The participating configurations were disentangled using the coincidences between the $^{22}$Mg core and the resulting gamma-rays. In the end the ANC for the $^{22}$Mg(0+)*proton configuration (the only one entering the inverse radiative proton capture reaction $^{22}$Mg(p,γ)$^{23}$Al) was extracted and used to evaluate the continuum contribution to the reaction rate. However, pertinent reaction rate (the NA result) was only obtained combining the result of this nuclear breakup experiment with that of the Coulomb breakup of the same projectile needed to evaluate the contribution of the resonant part. It is treated in the paper by A. Banu *et al*. and we refer the reader to it [30].

The uncertainty of 10-15% estimated for this method arose from a combination of experimental uncertainties but also from those of the calculations, using various approaches and effective nucleon-nucleon interactions. Question here are if this is 1) sufficient for NA and 2) if we have sufficient confidence in the types of calculations and parameters used? The answer to the first question is probably "yes, in most cases", while the second needs further work (this is the opinion of an experimentalist!). Which would mean that further work is needed to certify the reaction mechanism(s) of breakup – are those assumed the real ones? – and of the theoretical approaches used. To answer to the first question, we may need to use exclusive measurements, rather than inclusive ones, a task possible at the new RIB facilities.

### *D. The Trojan Horse Method*

This method is one of the most subtle among the indirect methods and is fully dedicated to nuclear astrophysics applications. It could also be called "the most direct of indirect methods in nuclear astrophysics". While it was initially proposed at about the same time as Coulomb dissociation by G. Baur [31], it was actually re-formulated and applied first at the end of the nineties (of the XX c.) by the Catania group lead by prof. C. Spitaleri. Many applications were made since by the same group of initiators, with experiments in Catania or in other laboratories. See e.g. [32-35], to just give a few examples. Theoretical developments were made in parallel, see [36, 37] and [20].

Briefly, the method works as follows. Instead of attempting the $A+x \rightarrow c+ C$ reaction at very low energies, experiment made difficult by the Coulomb barrier between charged nuclei $A$ and $x$, one does the experiment with 3-bodies in the final channel $A+a \rightarrow c+C + s$, at higher energies, above the Coulomb barrier. The projectile $a$ is chosen to be have a good cluster configuration $a=x +s$ in its ground state and the kinematic conditions are chosen such that the nucleus $x$ is moving slowly relative to the target $A$. This can stem from a combination of the projectile energy and the relative internal energy of $x$ and $s$ inside the compound $a$. While the relative energy $A$-$x$ is as low as in stellar reactions, $x$ is already beyond the Coulomb barrier and the reaction of interest takes place with larger probabilities. In the same time the nucleus (or nucleon) $s$ is a spectator. These are called quasi-free mechanism conditions. They must be fulfilled for the method to be applicable. With these kinematic conditions fulfilled, that is with the quasi-free mechanism present in reaction and with the cluster configuration of the projectile $a$ proven, one can make a direct connection between the triple differential cross section of the 3-body reaction measured and the cross section of the 2-body reaction at very low energies. The connection is easier to prove in the plane wave impulse approximation but is valid also in the distorted wave approximation. We send the reader to the detailed discussions in [20, 36, 37] and references therein.

To summarize, there are two main achievements of the THM:
- One can obtain data for very low energies, otherwise not accessible. In particular the behavior of the excitation functions close to $E_{cm}=0$ (or even below, see Ref. 34 for that). One can obtain the position of very low resonances and their widths and/or the contribution of sub-barrier resonances in cases where other methods fail [34].
- As the reaction A+x happens inside the barrier, it is a reaction between naked nuclei, with no screening from the electrons of the target and projectile, as is usually the case in direct laboratory measurements. There is no screening in stellar plasmas. Comparing the results of THM measurements with those of very low energy direct measurements one can obtain valuable and unique information about screening in nuclear reactions.

The method is useful in determining the behavior of the cross sections at very low energies but so far relies on normalizing the absolute values predicted to existing data at larger energies from direct measurements. The proportionality factor predicted by theory is not yet calculated. There are also discussions about the validity of the simpler plane wave approximation. Codes have been worked out to fit the data with multiple, overlapping resonances, they need to be further tested before being accepted by all parts in the discussions.

Recently the method could be applied for the first time for radioactive beams [35].

## *E. Spectroscopy of resonances*

Besides the continuum parts contributing to the reaction rate in stellar processes, contributions may occur from resonances. These resonances are meta-stable states in the compound nuclear system produced in reaction as an intermediate step in a two-steps process. The contribution of an isolated resonance at energy $E_r$ to the reaction rate of a stellar process at temperature T is given by [7]:

$$\langle \sigma \upsilon \rangle_{res} = \left( \frac{2\pi}{\mu kT} \right)^{3/2} \hbar^2 \omega\gamma \exp\left( -\frac{E_r}{kT} \right)$$

To evaluate the corresponding contributions to the reaction rates it is therefore, enough to determine the location of the resonances ($E_r$) and their resonance strengths ($\omega\gamma$) [7].

$\omega\gamma = (2J+1)/[(2j_i+1)(2j_o+1)] \, \Gamma_{in}\Gamma_{out}/\Gamma_{tot}$

The important resonances are located at very low energies, in the Gamow window, or around those energies (see for example [38]).

These quantities (actually the resonance strength is more than one parameter: we need the spin J of the state and the partial widths $\Gamma_{in}$ and $\Gamma_{out}$, with $\Gamma_{tot}=\Gamma_{in} + \Gamma_{out}$)) may be determined by studying the spectroscopic properties of the corresponding meta-stable state, populated through another, more convenient method than the low energy direct measurement (we use standard and obvious notations here). Both are equally important, as the dependence on the position of the resonance is exponential, and the resonance strength intervenes multiplicatively. In all cases the information on the quantum numbers (spin and parity $J^\pi$) for states located in the sensitive region is very important, because it tells if the meta-stable states in question can indeed be resonances that can contribute in the reaction studied. This is because at the low energies in stars only low partial waves (*s* or *p*) can typically contribute.

The types of measurements usable is obviously very large and diverse, and the list below is only schematic:
1. Transfer reactions
2. Gamma-ray spectroscopy
3. Beta-delayed proton emission
4. TTIK - Thick Target Inverse Kinematics scattering
5. Other spectroscopic methods.

It is beyond the scope and possibilities of this article to discuss each of them or give an exhaustive list of references. We would need to review virtually all nuclear physics spectroscopic methods to do that. Instead we only outline that all cases in experiments we attempt:
- To find the meta-stable states that may be resonances and determine their energy $E_r$
- To determine the spin and parity of the state, establish if the state found can be a contributing resonance in the reaction in stars
- Measure the partial widths that can lead to the determination of the resonance strength.

Without detailing further, we will just say that any experienced nuclear physicist knows that the latter is the most demanding of the steps!

*Decay spectroscopy.* We shall sketch only one of the types in the list above, *beta-delayed proton emission (βp)*, as it is newer, closer to us, and is becoming more frequently used due to the progress in the production of exotic nuclei in the new RIB facilities. This will also allow us to stress the improvements required in the experimental setups to obtain NA valuable information. The method works like this: instead of measuring radiative proton capture (p,γ) one can study the inverse of its first step, the proton decay of the same state. The decaying states are populated by beta-decay: in the same compound nucleus, states above the proton threshold are populated by β-decay, and then they decay emitting a proton. The method is applicable if the selection rules for (p,γ) and βp allow for the population of the same states (by the energy and spin-parity selection rules). One can determine that way the energy of the resonance, determine or restrict the spins and parity of the states involved and determine the branching ratios. This simple connection is schematically presented in figure 4 below From Ref. 21) for the case of the $^{22}$Na(p,γ)$^{23}$Mg radiative proton capture: we aim at populating and study states in the $^{23}$Mg daughter nucleus following the β-decay of $^{23}$Al.

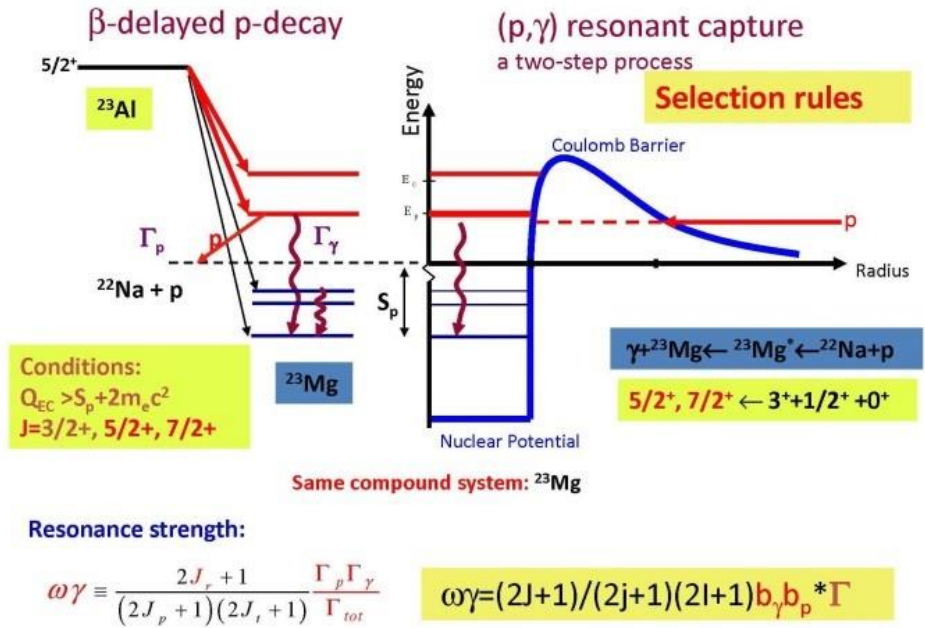

**Figure 5. Schematic correspondence of β-delayed proton-decay and resonant radiative proton capture.**

The selection rules allow that: s-wave radiative capture involves $J^\pi=5/2^+$ and $7/2^+$ states; beta-decay populates predominantly positive parity states with spins 3/2, 5/2 and 7/2. Figure 4 underlines that we need to locate the resonances and determine their properties (spin and parity and partial widths). Similar situations for other two proton capture reactions we studied through the decay of $^{27}$P and $^{31}$Cl, respectively. Measurements were done at the Cyclotron Institute of Texas A&M University using radioactive proton-rich nuclei produced and separated with the MARS recoil spectrometer. The short-lived radioactive species were produced in-flight (either $^{23}$Al, $^{27}$P, $^{31}$Cl, $^{20}$Mg etc., in most cases with purities 85% and up) and moving at 30-40 MeV/nucleon They were stopped and accumulated for about two lifetimes in a medium that was a detector (the implantation phase), then the beam was cut off and the protons from the β-delayed proton-decay were measured (the measurement phase). The detection medium was at first very thin silicon strip detectors (as thin as 65 and 45 μm) [39, 40], later a specially designed ionization chamber with micromegas gain amplifier [41]. While implanting the radioactive species in the very thin detectors mentioned above was an achievement per se because not only the RIB needed to be pure, but it needed to have a small spread of its incoming energy (not usual for those obtained from fragmentation or in-flight decay, but possible at MARS), the proton spectra in the region of interest (say 100-600 keV) were very much affected by the continuum background from the positrons emitted in the first step of the process [39]. The problem was more and more important toward lower proton energies, exactly those of interest in NA. Simultaneously, the proton-decay branching ratios become smaller at lower energies due to the barrier penetration factor in proton-decay. This overwhelming problem was diminished using gas as detection medium, as described in Ref. 41. The low amplitude signals from the decays in the detector that works in an ionizing chamber regime were then amplified with a micromegas systems of pads that also allowed for the diagnosis of the incoming beam of the implantation phase and the location of the decaying products in the second. Two devices ASTROBOX [41] and ASTROBOX2 [42] were realized and used in experiments with good results: beta- background free down to 80-100 keV and proton-decay branchings as low as $10^{-4}$ were obtained with these arrangements. We will skip the details in favor of sending the reader to the recent papers describing these experiments, the equipment and experimental methods involved, and their results. The method and the detection systems described can be used for other β-delayed charged-particle emission and worked even at radioactive beam rates of a few pps [43].

Recently a complex system based on same ideas was built at NSCL [44].

As a last point we want to stress what results from a careful inspection of the last equation shown in Fig. 4. The method allows for the identification of the location of resonances (Er) and for the determination of the proton and gamma decay branching, possibly of the spin and parity of the state(s), but does not allow the determination of the absolute value of the decay width(s) Γ, therefore of the evaluation of the absolute value of the resonance strength(s). The total decay width must be measured by other methods, for example by measuring the lifetime of the states through gamma-ray spectroscopy methods. This shows the complexity of the methods that must be used to get good nuclear data for NA.

## 5. Conclusions

As stated in the Introduction, the indirect methods of nuclear physics for astrophysics briefly presented would not be useful as standalone in nuclear astrophysics but need to be included in the whole environment that the problem of the origin of energy and of the elements in the Universe encompasses. As such a number of other directions of research must be considered. Therefore, stellar dynamics, nucleosynthesis modeling, observations (space-based telescopes, cosmochemistry, etc.) must be considered as part of the discussions. Discussions that need to involve members of what were, and in cases still are, considered different branches of physics.

The present review of existing indirect methods in nuclear astrophysics included the list of accepted methods, pointing to some specifics for each. Only brief assessments of problems with the accuracy of each indirect method, experimental and theoretical, stressing the importance of calculated absolute values. At points we specified the need for modern theories and codes, of better systematics, of tests of validity and of the parameters to use in calculations. A thorough review of the existing experimental meth-ods, equipment and specifics was not included here, as is beyond authors' abilities. Similarly, the new facilities, including RIB facilities, and their nuclear astrophysics programs were not discussed in this paper.

We can only point to discussions at the workshop on related topics and new directions: discussion and attention should be given also to the contribution of excited states to the processes in stellar plasma (the topic of the talk by A. Petrovici, see Ref. 45 and references therein.

Nuclear reactions in laser induced plasmas is becoming a hot topic in the last few years and are bound to become increasingly important soon after the first measurements were made in the newly available petawatt lasers [46, 47]. We dare to say that in the future laser induced plasmas will offer ways to evaluate experimentally the contribution of the excited states to nuclear reaction rates in stars, while currently only theoretical predictions are being made [45].

The problem of the equation of state of nuclear matter, crucial for the connection between nuclear physics and neutron stars, was not attempted here (and at the workshop). Nor did have a large, thorough, coverage the problem of the structure of neutron-rich nuclei on the path of the r-process. There are and shall be topics for dedicated meetings and papers.

## *Acknowledgements*

One of us (LT) acknowledges the discussion had with profs. C. Bertulani, A. Bonaccorso, Zs. Fülop and T. Motobayashi, co-organisers of the ECT* workshop that lead to the present paper. His stay and the workshop were supported in part by ECT* and by the project ENSAR2. ENSAR2 has received funding from the European Union's Horizon 2020 research and innovation programme under grant agreement No. 654002. This work was supported in part by the grant PNIII/P5/P5.2 no. 02/FAIR-RO from the Romanian Ministry of Research and Innovation through its IFA funding agency.